\documentclass[a4paper,11pt]{article}
\pdfoutput=1 
\usepackage{jheppub} 

\usepackage[T1]{fontenc} 

\RequirePackage[displaymath]{lineno} 
\usepackage{epstopdf}
\usepackage{epsfig}
\usepackage{graphicx}
\usepackage{dcolumn}
\usepackage{bm}
\usepackage{ltablex,booktabs}
\usepackage{overpic}
\usepackage{subfigure}
\usepackage{float}
\usepackage{color}
\usepackage{amsmath}
\usepackage{mathcomp}
\usepackage{mathrsfs}
\usepackage{multirow}
\usepackage{rotating}
\usepackage{amssymb}
\usepackage{gensymb}
\usepackage{amsmath}
\usepackage{tabularx}
\usepackage{threeparttable}

\title{\boldmath Observation of $\psi(3686) \to \Omega^- K^+ \bar{\Xi}^0 $+c.c.}

\collaborationImg{\includegraphics[width=0.15\textwidth, angle=90]{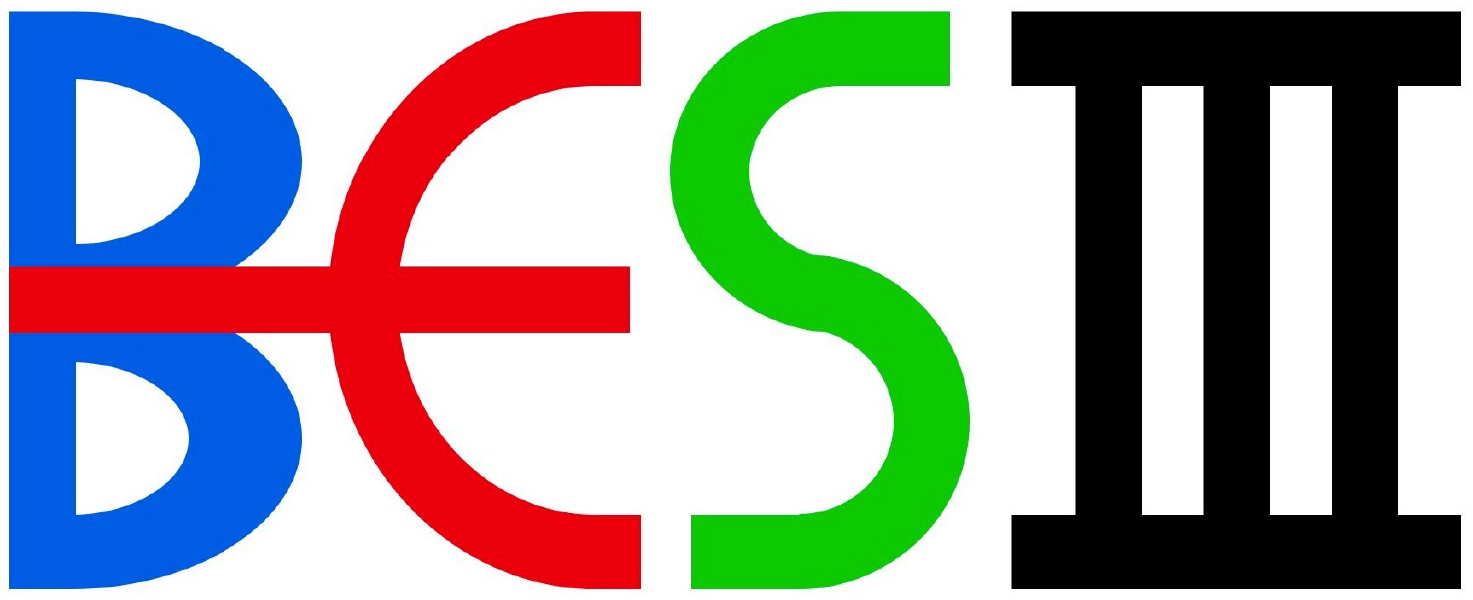}}
\collaboration{BESIII Collaboration}

\author{
M.~Ablikim$^{1}$, M.~N.~Achasov$^{4,d}$, P.~Adlarson$^{75}$, O.~Afedulidis$^{3}$, X.~C.~Ai$^{80}$, R.~Aliberti$^{35}$, A.~Amoroso$^{74A,74C}$, Q.~An$^{71,58,b}$, Y.~Bai$^{57}$, O.~Bakina$^{36}$
% ...
\\
\vspace{0.2cm}
(BESIII Collaboration)\\
\vspace{0.2cm} {\it
$^{1}$ Institute of High Energy Physics, Beijing 100049, People's Republic of China\\
% ...
$^{3}$ Bochum  Ruhr-University, D-44780 Bochum, Germany\\
$^{4}$ Budker Institute of Nuclear Physics SB RAS (BINP), Novosibirsk 630090, Russia\\
% ...
$^{35}$ Johannes Gutenberg University of Mainz, Johann-Joachim-Becher-Weg 45, D-55099 Mainz, Germany\\
$^{36}$ Joint Institute for Nuclear Research, 141980 Dubna, Moscow region, Russia\\
% ...
$^{57}$ Southeast University, Nanjing 211100, People's Republic of China\\
$^{58}$ State Key Laboratory of Particle Detection and Electronics, Beijing 100049, Hefei 230026, People's Republic of China\\
% ...
$^{71}$ University of Science and Technology of China, Hefei 230026, People's Republic of China\\
% ...
$^{74}$ University of Turin and INFN, (A)University of Turin, I-10125, Turin, Italy; (B)University of Eastern Piedmont, I-15121, Alessandria, Italy; (C)INFN, I-10125, Turin, Italy\\
$^{75}$ Uppsala University, Box 516, SE-75120 Uppsala, Sweden\\
% ...
$^{80}$ Zhengzhou University, Zhengzhou 450001, People's Republic of China\\

\vspace{0.2cm}
$^{b}$ Deceased\\
% ...
$^{d}$ Also at the Novosibirsk State University, Novosibirsk, 630090, Russia\\
% ...

}
}

\abstract{ 
Using $(27.12 \pm 0.14) \times 10^{8}$ $\psi(3686)$ events collected with the BESIII detector at BEPCII, the decay of $\psi(3686) \to \Omega^- K^+ \bar{\Xi}^0 +c.c.$ is observed for the first time. The branching fraction of this decay is measured to be $\mathcal{B}_{\psi(3686) \to \Omega^- K^+ \bar{\Xi}^0 +c.c.}=(2.78 \pm 0.40 \pm 0.18 ) \times 10^{-6}$,
where the first uncertainty is statistical and the second is systematic. 
Possible baryon excited states are searched for in this decay, but no evident intermediate state is observed with the current sample size.

}
\arxivnumber{}

\begin{document}
\maketitle
\flushbottom

\section{Introduction}
\label{sec:introduction}

The discovery of the $J/\psi$ and other  $c\bar{c}$ bound states had a great impact on the development of the theory of strong interaction within the Standard Model (SM) ~\cite{E598:1974sol,PhysRevLett.33.1406}. These states are multi-scale systems
that probe a wide span of energy regimes in quantum chromodynamics~(QCD): from the hard region, where expansions in the coupling constant are legitimate, to the low-energy region, where nonperturbative effects dominate~\cite{Asner:2008nq}. Heavy quark-antiquark states remain an ideal laboratory where non-perturbative QCD and its interplay with perturbative QCD can be tested in a controlled framework. 

The observation and study of a decay with three pairs of $s\bar{s}$ in the final state will expand our knowledge of the decay mechanism of charmonium and has the potential to improve of understanding of QCD~\cite{BESIII:2020lkm,BESIII:2023olq}. 
Experimental studies of hadronic decays of charmonium states provide important information for investigating many topics involving the strong interaction, such as 
the color octet and singlet contributions, the violation of helicity conservation, and SU(3) flavor symmetry breaking effects~\cite{Asner:2008nq,Rybicki:2009zza}.

Compared to the two-body final state, the theoretical analysis relevant to three-body decays of charmoniums is more difficult and the available experimental results are rather limited at present~\cite{PDG}. 
Recently, theoretical interest in final states containing baryons has been revived, stimulated by recent experimental discoveries, especially the phenomena of baryon-anitbaryon invariant mass enhancements near threshold~\cite{BES:2003aic,BES:2004fgd,BESIII:2012koo}.


The study of baryon spectroscopy played an important role in the development of the quark model and QCD~\cite{PDG,Capstick:2000qj}, although our knowledge of this subject is still limited.
Due to the small production cross sections and the complicated topology of
the final states, only a few $\Xi^*$ and $\Omega^*$ states have been observed to date, and many of them lack a spin-parity determination. Until now, the most useful measurements have come from diffractive $K^-p$ interactions~\cite{PDG,Aston:1987bb}.

In this paper, the first observation of the decay $\psi(3686) \to \Omega^- K^+ \bar{\Xi}^0 +c.c.$ is reported and the corresponding
branching fraction is measured using $(27.12 \pm 0.14) \times 10^{8}$
$\psi(3686)$ events~\cite{BESIII:2024lks} collected with the BESIII detector.
In addition, possible baryon excited states are also searched for in this decay. Throughout this paper, the charge conjugation decay mode
is always implied.

\section{BESIII detector and Monte Carlo simulation}
\label{sec:BESIII and MC}
The BESIII detector~\cite{BESIII:2009fln, BESIII:2020nme} records $e^+ e^-$ collisions provided by the BEPCII storage ring~\cite{Yu:2016cof}, which operates with a peak luminosity of $1\times 10^{33}\; \text{cm}^{-2}\text{s}^{-1}$ 
in the  center-of-mass energy range from 2.00 to 4.95 GeV. 
The cylindrical core of the BESIII detector covers 93\% of the full solid angle and consists of a helium-based multilayer drift chamber (MDC), 
a plastic scintillator time-of-flight system (TOF), and a CsI (Tl) electromagnetic calorimeter (EMC), 
which are all enclosed in a superconducting solenoidal magnet providing a 1.0~T magnetic field~\cite{Huang:2022wuo}. 
The solenoid is supported by an octagonal flux-return yoke with resistive plate counter muon identification modules interleaved with steel. 
The charged-particle momentum resolution at 1~GeV/$c$ is 0.5\%, and the  d$E/$d$x$ resolution is 6\% for the electrons from Bhabha scattering at 1~GeV. 
The EMC measures photon energies with a resolution of 2.5\% (5\%) at 1~GeV in the barrel (end-cap) region. 
The time resolution of the TOF barrel part is 68~ps, while that of the end-cap part was 110~ps. 
The end-cap TOF system was upgraded in 2015 using multi-gap resistive plate chamber technology, providing a time resolution of 60~ps~\cite{Li:2017eToF, Guo:2017eToF, Cao:2020ibk}.

Monte Carlo (MC) simulated data samples produced with a {\sc geant4}~\cite{GEANT4:2002zbu} based software package, which includes the geometric description of the BESIII detector and the
detector response, are used to optimize the event selection criteria, estimate the signal efficiency and background level. 
The simulation models the beam-energy spread and initial-state radiation in the $e^+e^-$ annihilation using the generator 
{\sc kkmc}~\cite{Jadach:2000ir}.
The inclusive MC sample includes the production of the $\psi(3686)$ resonance, the initial-state radiation production of the $J/\psi$ meson, and the continuum processes
incorporated in {\sc kkmc}~\cite{Jadach:2000ir}. Particle decays are generated by {\sc evtgen}~\cite{Lange:2001uf, Ping:2008zz}
for the known decay modes with branching fractions taken from the Particle Data Group~\cite{PDG} and {\sc lundcharm}~\cite{Chen:2000tv, Yang:2014vra} for the remaining unknown ones. 
Final-state radiation from charged final-state particles is included using the {\sc photos} package~\cite{Richter-Was:1992hxq}.
To determine the detection efficiency, a signal MC sample of the whole decay chain of $\psi(3686) \to \Omega^- K^+ \bar{\Xi}^0, \Omega^- \to \Lambda(\to p \pi^-) K^-$ is generated uniformly in phase space~(PHSP), along with generic $\bar{\Xi}^0$ decays.
An inclusive $\psi(3686)$ MC sample, consisting of $27.12 \times 10^8$ events, is used to estimate potential backgrounds. The data sample collected at the center-of-mass energy of 3.650, 3.682, 3.773 GeV and 9 energy points from 3.58 to 3.71 GeV,
corresponding to total integrated luminosities of 410~pb$^{-1}$, 404~pb$^{-1}$, 7.93~fb$^{-1}$ and 503~pb$^{-1}$, respectively, are used to estimate the contamination from the continuum processes.

\section{Event selection}
\label{sec:event selection}
The cascade decay of interest is $\psi(3686) \to \Omega^- K^+\bar{\Xi}^0$,~with $\Omega^- \to \Lambda K^-$, $\Lambda \to p \pi^- $.  
As the full reconstruction method suffers from low detection efficiency, a partial-reconstruction strategy is applied,  in which only the $\Omega^- K^+$ is reconstructed, with no attempt made to identify the $\bar{\Xi}^0$. 
The charged tracks in the MDC are required to have a polar angle $\theta$ with respect to the beam direction within the MDC acceptance
$|\!\cos\theta|<0.93$. 
For each charged track, particle identification (PID) is performed, combining measurements of the d$E/$d$x$ in the MDC and the flight time in the TOF to form particle identification~(PID) confidence levels~$CL_{h}$ for each hadron $h$ hypothesis.
The charged tracks are identified as protons with the requirement of $CL_{p}>CL_{K}$, $CL_{p}>CL_{\pi}$ and $CL_{p}>0.001$, and kaons with $CL_{K}>CL_{\pi}$ and $CL_{K}>0$. The remaining charged tracks are assigned to be pions.
If there is more than one $K^+$ candidate, the $K^+$ with the highest $CL_{K}$ is assumed to be from the interaction point (IP), $i.e.$ the $K^+$ is further required to have a distance of closest approach to the IP less than 10~cm along the $z$-axis and less than 1~cm in the transverse plane. 

\begin{figure}[htbp]
	\begin{center}
        \mbox{
            \put(-230, 10){
                \begin{overpic}[width = 0.5\linewidth]{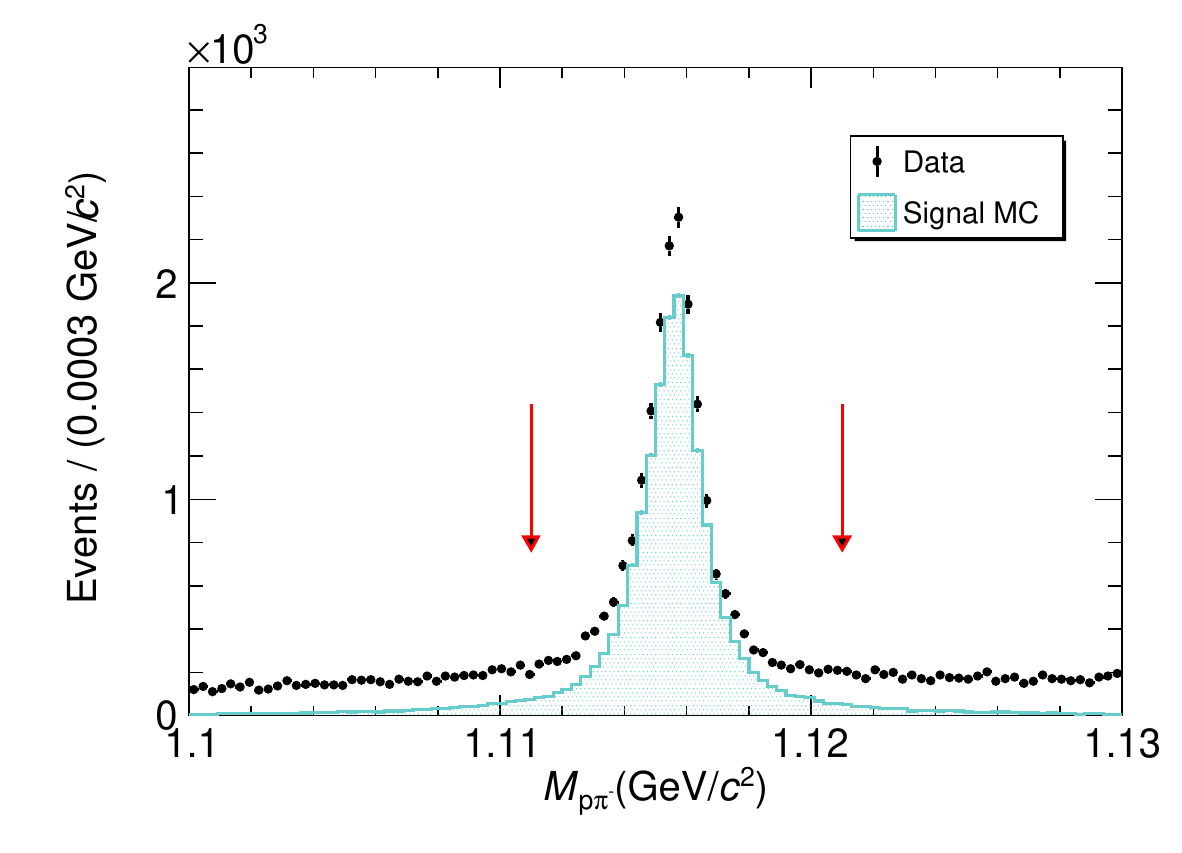}
                \end{overpic}}
            \put(-10, 10){
                \begin{overpic}[width = 0.5\linewidth]{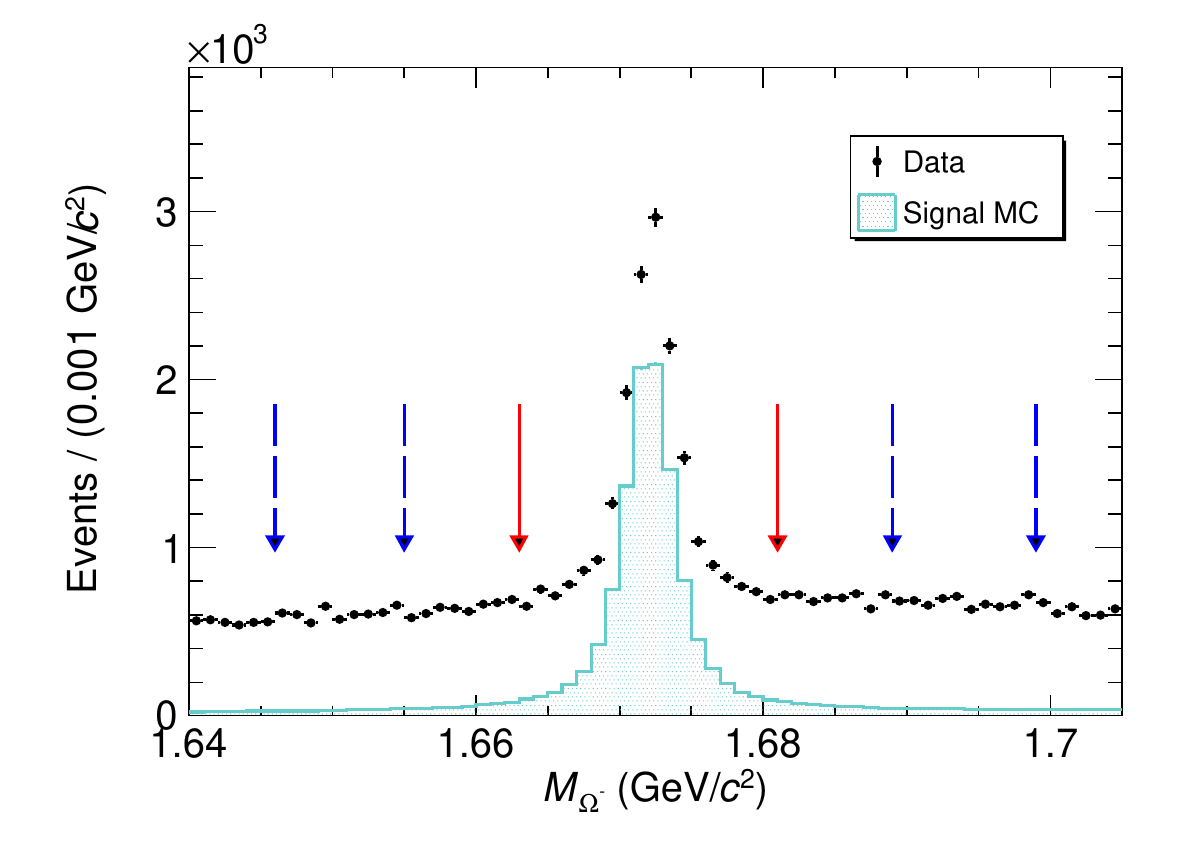}
                \end{overpic}}
            \put(-195, 165) { (a)}
            \put(30, 165)   { (b)}
        }
	\end{center}
	\caption{
The distributions of $M_{p \pi^-}$ and $M_{\Omega^-}$. 
The red arrows show the signal region, and the blue dashed arrows show the sideband regions. 
	}
	\label{Lambda and Omega}
\end{figure}

\begin{figure*}[htbp]
    \begin{center}
        \mbox{
            \put(-230, 10){ \begin{overpic}[width = 0.5\linewidth]{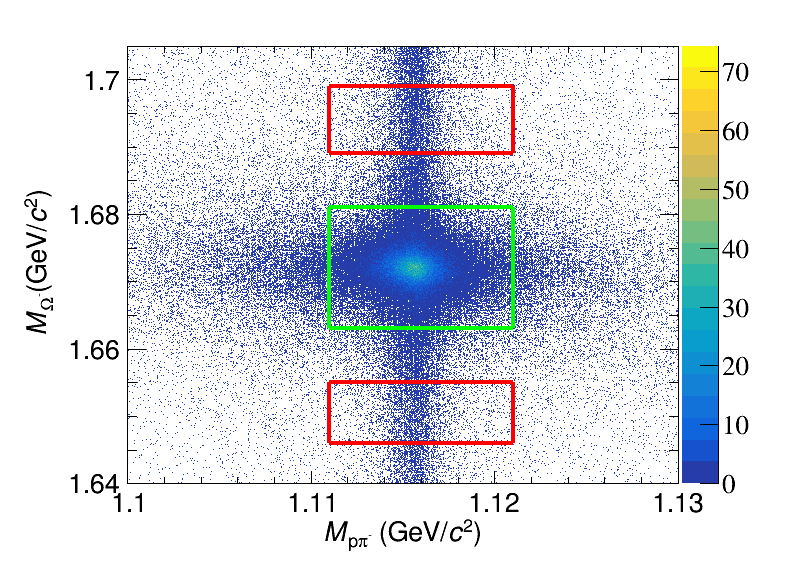}
            \end{overpic}}
                
            \put(-10, 10){\begin{overpic}[width = 0.5\linewidth]{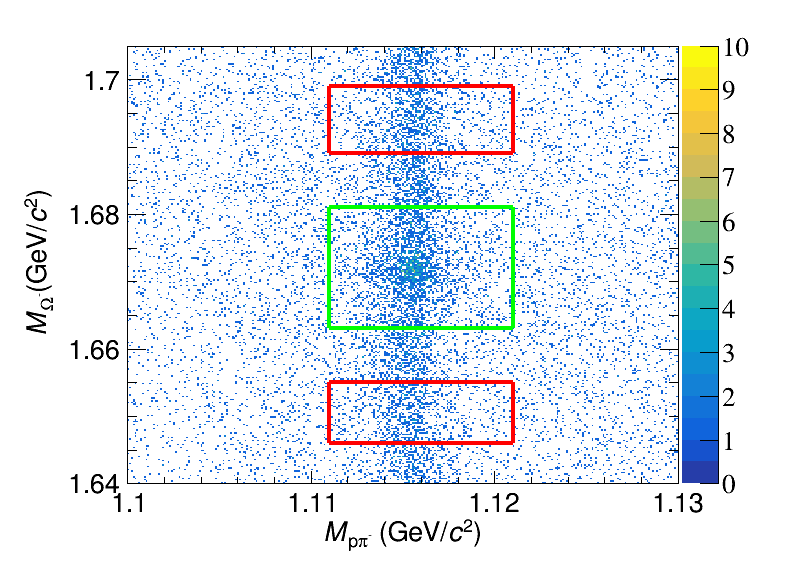}
            \end{overpic}}
            
            \put(-195, 165) {$(a)$}
            \put(30, 165) { $(b)$}
        }
    \end{center}
    \caption{
        The 2-D distributions of $M_{p\pi^-}$  and $M_{\Omega^-}$ for signal MC sample (a) and data (b). The green box denotes the signal region and the red boxes denote the sideband regions.
    }
    \label{fig:Lam vs Omega}
\end{figure*}

\begin{figure}[htpb]
    \centering
    \includegraphics[width=0.6\textwidth]{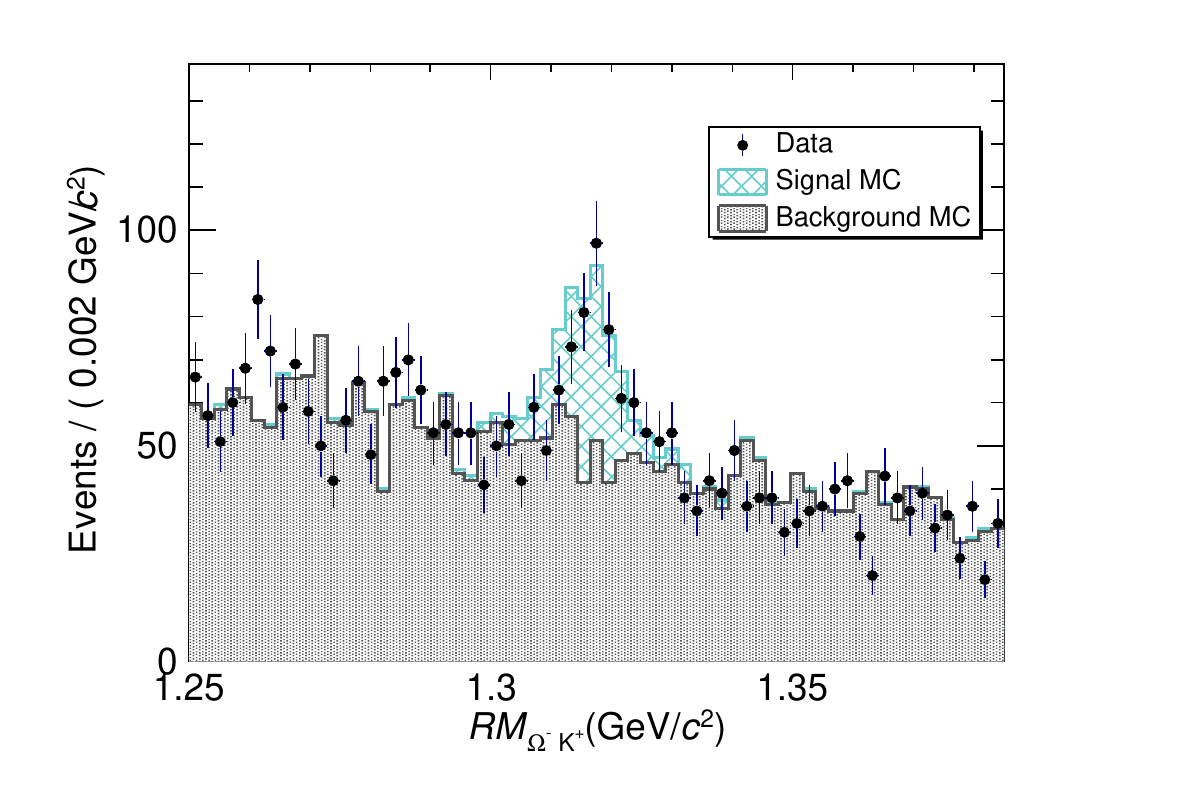}
    \caption{The distribution of $RM_{\Omega^- K^{+}}$.}
    \label{fig:RMOmegaK}
\end{figure}

$\Lambda$ candidates are reconstructed using common vertex fits~\cite{Xu:2009zzg} on $p \pi^-$ pairs with the requirement $\chi^{2}<200$. If there is more than one $p \pi^-$ combination, the pair corresponding to the minimum $\chi^2$ from the vertex fit is retained.  
The $p \pi^-$ invariant mass ($M_{p \pi^-}$) must be within the $\Lambda$ signal region, $M_{p \pi^-} \in [1.111, 1.121]$~GeV/$c^2$, as shown in Fig.~\ref{Lambda and Omega} (a).
The $\Omega^-$ decay is reconstructed with a $\Lambda$ candidate and a $K^-$ by implementing another common vertex fit for which $\chi^2<200$ is again required. 
The reduced mass for the $\Omega^-$ candidates, $M_{\Omega^-}=M_{\Lambda K^-}-M_{p \pi^-}+M_{\Lambda}^{\rm PDG}$, is used to improve the mass resolution of  the $\Omega^-$ candidate, where $M_{\Lambda}^{\rm PDG}$ is the known mass of the $\Lambda$ baryon~\cite{PDG}. If there is more than one $\Omega^-$ candidate, the one with the minimum $|M_{\Omega^-}-M_{\Omega^-}^{\rm PDG}|$ is chosen, where $M_{\Omega^-}^{\rm PDG}$ is the known mass of the $\Omega^-$~\cite{PDG}.
Based on the MC study, the probability of occurrence of multiple $\Omega^-$ candidates is about 1.7\%. With our method of selecting the best $\Omega^-$ candidate, only less than four percent of them would select the incorrect $\Omega^-$ candidate,
which means the potential bias of best $\Omega^-$ candidate selection is negligible.
The distribution of $M_{\Omega^-}$ is shown in Fig.~\ref{Lambda and Omega} (b). The signal region defined as $M_{\Omega^-}$ $\in$ $[1.663, 1.681]$~GeV/$c^2$, corresponding to six times the mass resolution, is imposed to select $\Omega^-$ candidates. The sideband regions defined as $M_{\Omega^-} \in ([1.646, 1.655]$ $\cup$ $[1.689, 1.699])$~GeV/$c^2$ are used to estimate the background.
The two-dimensional~(2-D) distributions of $M_{p\pi^-}$ versus $M_{\Omega^-}$ for signal MC sample and data are shown in Fig.~\ref{fig:Lam vs Omega}.
Signal events manifest themselves  through a $\bar{\Xi}^0$ peak in the sprectrum of the invaritant mass recoiling against the $\Omega^- K^+$ pair~($RM_{\Omega^- K^+}$).
The $RM_{\Omega^- K^+}$ spectrum for MC sample and data are shown in Fig.~\ref{fig:RMOmegaK}. 
The $\bar{\Xi}^0$ signal region is defined as $RM_{\Omega^- K^+}\in$[1.282, 1.352] GeV/$c^2$, corresponding to approximately $\pm 3\sigma$ around the nominal $\bar{\Xi}^0$ mass, where $\sigma$ is the mass resolution of $RM_{\Omega^- K^+}$ from the signal MC sample.

\section{Detection efficiency determination}
\label{sec:detection efficiency}

\begin{figure}[htpb]
    \centering
    \includegraphics[width=0.6\textwidth]{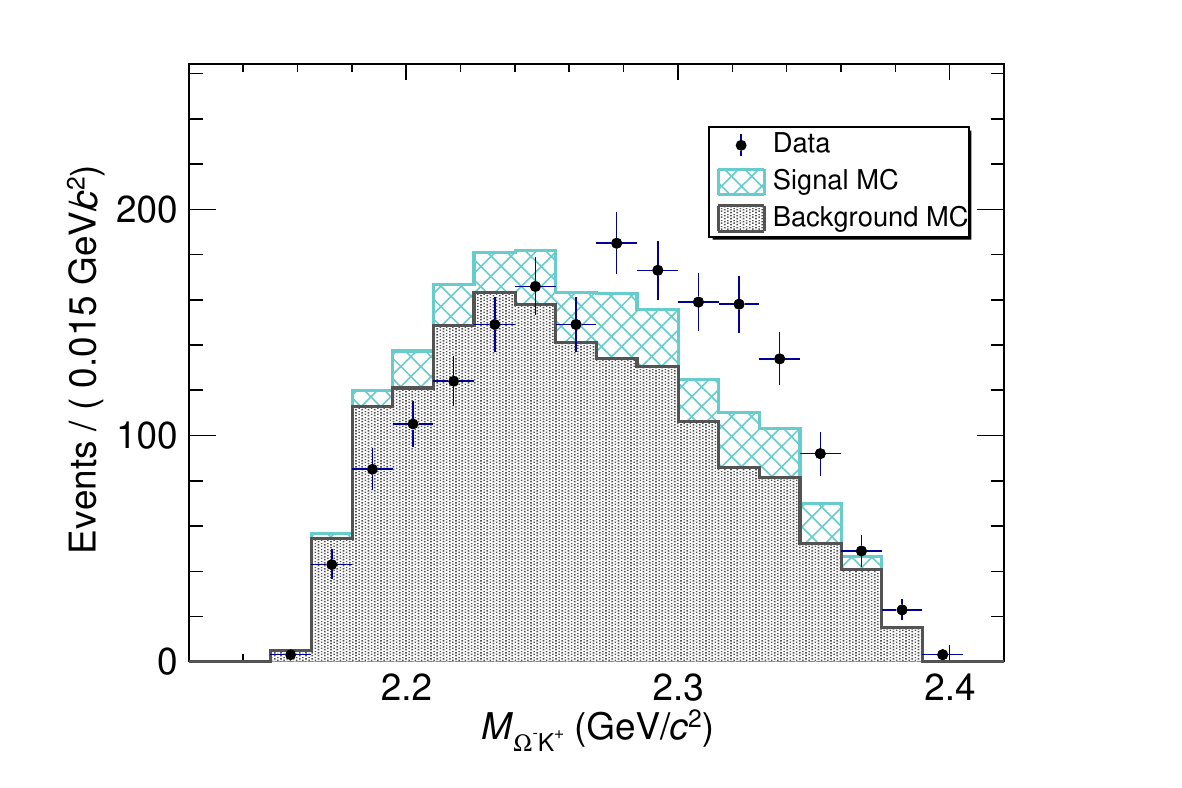}
    \caption{The distribution of $M_{\Omega^-K^+}$.}
    \label{fig:two-body mass}
\end{figure}

The detection efficiency is determined with MC simulation.
Thus, it is necessary to assess the potential impact of intermediate states.
Based on the known excited states of $\Omega^-$ and $\Xi^0$~\cite{PDG}, under the limitation of the PHSP, the only possible excited states in this channel is $\Xi(2250)^0(\to \Omega^- K^+)$.
As a further test to ensure we are not sensitive to intermediate resonances, we combine the signal MC sample with the inclusive MC sample, based on the measured branching fraction of this decay~(which can be found in Sec.~\ref{sec:BF}), and verify its consistency with the data.
The distributions of $M_{\Omega^-K^+}$ between data and MC simulations in Fig.~\ref{fig:two-body mass} 
show acceptable agreement, 
and no intermediate state is evident in the data sample.


\section{Background study}
\label{sec:background study}
The $\psi(3686)$ inclusive MC sample is analysed with a generic event-type examination tool, TopoAna~\cite{Zhou:2020ksj}, to identify potential backgrounds.
Further studies are performed of the surviving events in the $\bar{\Xi}^0$ signal region from the inclusive MC sample, and the events in the $\Omega^-$ mass sideband regions from data, respectively. 
These investigations indicate that there is no significant source of peaking background in the $RM_{\Omega^-K^+}$ spectrum. 

To investigate the contamination from continuum processes 
\cite{BESIII:2024lks}, the same selection criteria are applied to the control samples introduced in Sec.~\ref{sec:BESIII and MC}. Peaking background events from the data sample at center-of-mass energy of 3.773 GeV are observed in Fig.~\ref{fig:RMOmegaK_fit} (b) from Sec.~\ref{sec:BF}, while the background  from the other continuum data samples are negligible. 

\section{Signal yield and BF}
\label{sec:BF}
\begin{figure*}[htbp]
    \begin{center}
        \mbox{
            \put(-230, 10)
            {
                \begin{overpic}[width = 0.5\linewidth]{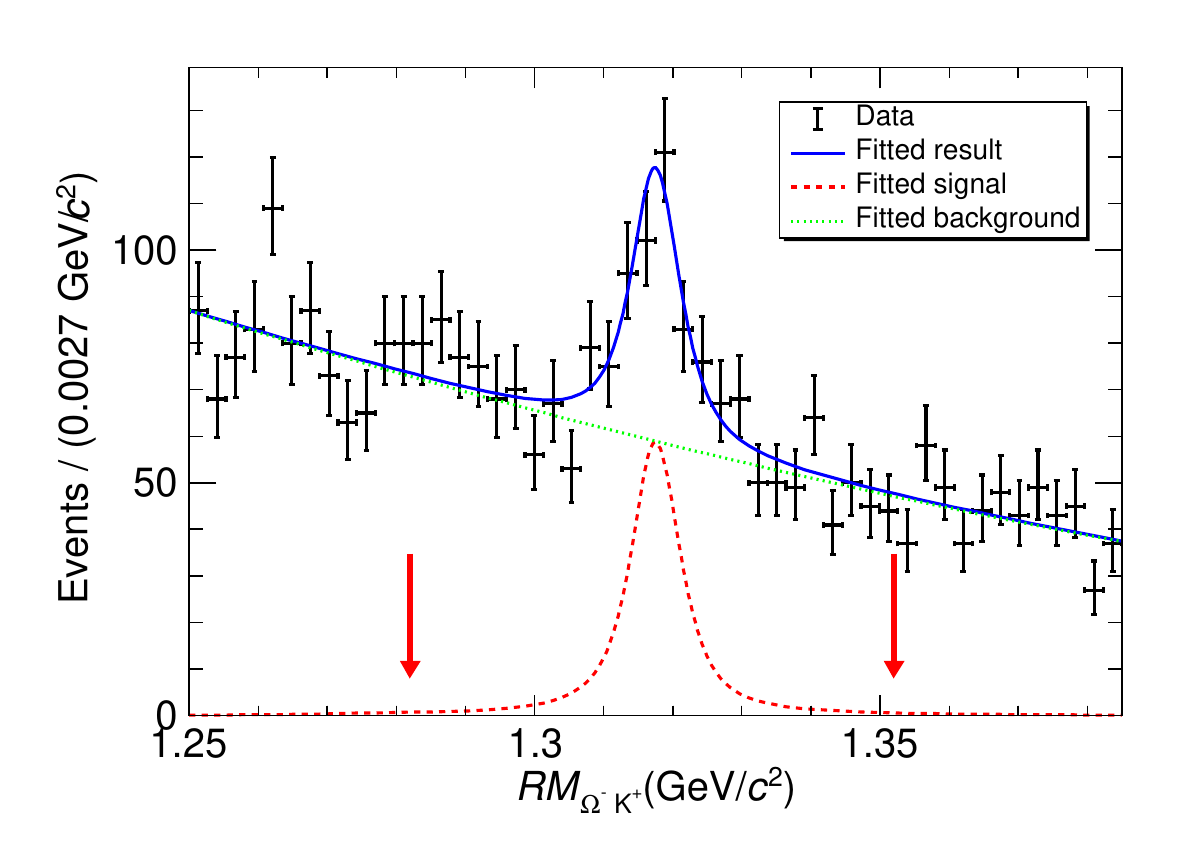}
                \end{overpic}
            }
            \put(-10, 10)
            {
                \begin{overpic}[width = 0.5\linewidth]{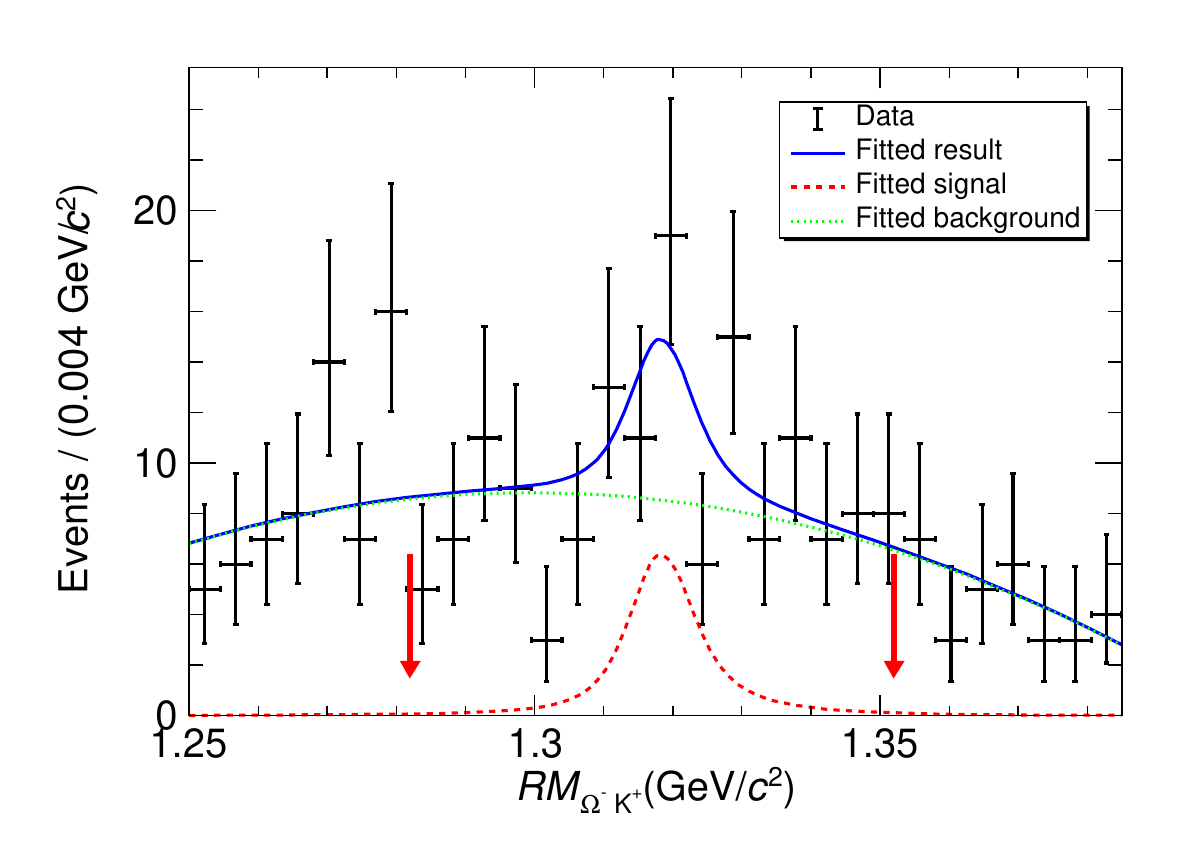}
                \end{overpic}
            }
            \put(-195, 160) { $(a)$}
            \put(30, 160)  {$(b)$}
        }
    \end{center}
    \caption{
     Fits to the $RM_{\Omega^- K^+}$ distributions of the accepted candidates in $\psi(3686)$ data (a) and the continuum data at 3.773 GeV (b).
The red arrows mark the $\bar{\Xi}^0$ signal region.
    }
    \label{fig:RMOmegaK_fit}
\end{figure*}

To determine the signal yield, an unbinned maximum-likelihood fit is performed on the $RM_{\Omega^- K^+}$ distribution.
In the fit, the signal shape is described by the signal MC simulated shape convolved with a Gaussian function with free parameters, where the Gaussian function is used to compensate for the difference in mass resolution between data and MC simulation. The background shape is described by a second-order Chebyshev polynomial. The fit result is shown in Fig.~\ref{fig:RMOmegaK_fit} (a). The signal yield from $\psi(3686)$ data in the signal region is determined to be $N_\textrm{obs.} = 250 \pm 35$.
The statistical significance of the $\Bar{\Xi}^0$ signal is $8.3\sigma$,
which is determined from the change in the log-likelihood values and the corresponding change in the number of degrees of freedom with and without including the signal contribution in the fit. 

The same fit procedure is performed on the continuum data at 3.773 GeV~\cite{Ke:2023qzc}. The parameters of the Gaussian function used for the convolution here are also floating. The result of the fit to the $RM_{\Omega^- K^+}$ distribution is shown in Fig.~\ref{fig:RMOmegaK_fit} (b). The number of continuum background events fitted as signal in this sample is $N_\textrm{cont.}=21\pm11$, as shown in Table~\ref{tab:QED background}.
A scale factor $f_c$ is defined as the ratio of the normalized number of continuum background events ($N_\textrm{QED}$) to that in the data at 3.773 GeV, so that $N_\textrm{QED} = f_c N_\textrm{cont.}$, where

\begin{equation}
    f_\textrm{c} = \frac{\mathcal{L}_{\psi(3686)}}{\mathcal{L}_\textrm{cont.}} \cdot \frac{s_{\textrm{cont.}}^n}{s_{\psi(3686)}^n} \cdot \frac{\epsilon_{\psi(3686)}}{\epsilon_\textrm{cont.}}.
    \label{eq:f_c}
\end{equation}
Here, $\mathcal{L}$~\cite{BESIII:2024lks, Ablikim:2013ntc}, $s$, and $\epsilon$ refer to the integrated luminosity of data samples, the square of the center-of-mass energy, and the detection efficiency at the two center-of-mass energies, respectively. 
The index number $n$ is assumed to be 1 for the baseline result, which corresponds to a $1/s$ dependence for the input cross sections.
The impact of this assumption for $n$ will be considered as a source of systematic uncertainty. The scale factor is calculated to be 0.36 and $N_\textrm{QED}$ is $8\pm 4$. All relevant numbers are listed in Table~\ref{tab:QED background}.

\begin{table}[htbp]
    \centering
    \caption{The corresponding physical quantities in $f_\textrm{c}$ factor and the estimated values of $f_\textrm{c}$ and $N_\textrm{QED}$.}
    \begin{tabular}{l  c c | c c c}
    \hline \hline
    $\sqrt{s}$(GeV) & $\mathcal{{L}}(\textrm{fb}^{-1})$     & $\epsilon(\%)$ &     $N_\textrm{cont.}$ &  $f_\textrm{c}$ & $N_\textrm{QED}$\\
    \hline
    3.686 & 4.08 & 7.43 & \multirow{2}{*}{$21\pm11$} & \multirow{2}{*}{ 0.36} & \multirow{2}{*}{$8\pm4$ } \\
    3.773 & 7.93 & 11.09  & & &\\
    \hline \hline
    \end{tabular}
    \label{tab:QED background}
\end{table}


Due to the limited sample size of the data  taken in the vicinity of the $\psi(3686)$, the interference phase between the $\psi(3686)$ decay and the continuum production cannot be determined.
Furthermore, the signal can be described well by the pure signal MC ($\psi(3686) \to \Omega^- K^+ \bar{\Xi}^0$) shape convolving with a Gaussian function as shown in Fig.~\ref{fig:RMOmegaK_fit} (a), indicating that the potential interference is a subleading effect.
Thus in this analysis, we do not consider the interference effect between the $\psi(3686) \to \Omega^- K^+ \bar{\Xi}^0$ decay and the continuum production $e^{+}e^{-} \to \Omega^- K^+ \bar{\Xi}^0$.

The branching fraction of the $\psi(3686) \to \Omega^- K^+ \bar{\Xi}^0$ decay is calculated as
\begin{footnotesize}
    \begin{equation}
        \begin{aligned}
         \mathcal{B}_{\psi(3686) \to \Omega^- K^+ \bar{\Xi}^0 +c.c.} = \frac{N_\textrm{obs.} - N_\textrm{QED}} {N_{\psi(3686)} \cdot \mathcal{B}_{\Omega^- \to \Lambda K^-} \cdot \mathcal{B}_{\Lambda \to p^+ \pi^-} \cdot \epsilon},
         \end{aligned}
    \end{equation}
\end{footnotesize}
\setlength{\parindent}{0pt}

where $N_\textrm{obs.}-N_\textrm{QED}=242\pm35$ is the net number of signal events, $N_{\psi(3686)}$ is the total number of $\psi(3686)$ events~\cite{BESIII:2024lks}, and $\epsilon=7.43\%$ is the detection efficiency. $\mathcal{B}_{\Omega^- \to \Lambda K^-}$ and $\mathcal{B}_{\Lambda \to p^+ \pi^-}$ are the branching fractions for $\Omega^- \to \Lambda K^-$, and $\Lambda \to p^+ \pi^-$ decays, respectively, cited from the PDG~\cite{PDG}. With these inputs, the branching fraction of $\psi(3686) \to \Omega^- K^+ \bar{\Xi}^0 +c.c.$ is determined to be $(2.78\pm0.40)\times10^{-6}$.


\section{Systematic uncertainty}
\label{sec:systematic uncertainty}
The systematic uncertainties in the  $\mathcal{B}_{\psi(3686) \to \Omega^- K^+ \bar{\Xi}^0}$ measurement include contributions associated with the kaon-tracking efficiency, PID, $\Lambda$ reconstruction, the requirement on $M_{\Omega^-}$, signal and background shapes, 
$f_\textrm{c}$ factor, MC generator, the sample size of the MC sample, the input branching fractions~\cite{PDG}, and the total number of $\psi(3686)$ events~\cite{BESIII:2024lks}. 

\hspace{0.7cm}The systematic uncertainties arising from the knowledge of the kaon-tracking and PID efficiencies are studied with the well understood decay of $e^{+} e^{-} \to K^{+} K^{-} $, and both assigned as 1.0\% per track~\cite{BESIII:2018ldc}. 
The systematic uncertainty associated with the $\Lambda$-reconstruction efficiency includes  effects from the tracking (PID) efficiencies for proton and pion, and the requirement on $M_{p\pi^-}$. 
This uncertainty is estimated with a control sample of $J/\psi \to p K^- \bar{\Lambda} + c.c.$ decays~\cite{BESIII:2023drj}.
The momentum-dependent ratios of the $\Lambda$ reconstruction efficiencies between data and MC simulation are used to re-weight the MC sample. The difference between the baseline detection efficiency and that obtained after re-weighting, 4.1\%, is taken as the systematic uncertainty.



\hspace{0.7cm}The systematic uncertainty associated with the requirement on $M_{\Omega^-}$ is studied with a control sample of $\psi(3686) \to \Omega^- \bar{\Omega}^{+}$ events, where $\Omega^-$ is fully reconstructed with $\Omega^-\to \Lambda K^-$, $\Lambda \to p\pi^-$. Details on the event selection can be found in Ref.~\cite{BESIII:2023ldd}. 
The signal yield is obtained by fitting the recoiling mass against the $\Omega^-$~($RM_{\Omega^-}$). 
In the baseline analysis, the requirement of $M_{\Omega^-}\in[1.663, 1.681]$ GeV/$c^2$ is applied,
which is about $\pm 3\sigma$ around the known $\Omega^-$ mass. 
We change the $\Omega^-$ mass window to be $[1.643, 1.700]$~GeV/$c^2$, which contains almost all the signal events and retains a low background level.
The change in the efficiency difference between data and MC simulation is taken as the systematic uncertainty, which is 0.6\%.


\hspace{0.7cm}The systematic uncertainty associated with the knowledge of the signal shape is caused by detection resolution effects in data, and is estimated by
changing the MC shape convolved with a Gaussian function to the MC shape convolved with a double-Gaussian function. 
The relative difference in the branching fraction, 0.4\%, is assigned as the uncertainty.
The systematic uncertainty associated with the background shape is estimated by changing the background shape from a second-order Chebyshev polynomial to a third-order Chebyshev polynomial. The resulting difference to the
original branching fraction, 0.4\%, is assigned as the systematic uncertainty.

\hspace{0.7cm}The systematic uncertainty related to the scale factor $f_\textrm{c}$ is estimated by changing $n$ from 1 to 2 and 3. The resulting largest difference to the original branching fraction, 0.3\%, is assigned as the systematic uncertainty.
  
\hspace{0.7cm}Similar to Ref.~\cite{BESIII:2023ldd}, a event-by-event weighting method is used to study the systematic uncertainty related to the MC generator. 
The events of the signal MC sample are weighted in two dimensions according to the momentum distribution of $K^+$ and $\Omega^-$ in data.
The deviation between the nominal detection efficiency and that obtained after re-weighting, 3.9\%, is taken as the systematic uncertainty.
  
\hspace{0.7cm}The systematic uncertainty arising from the limited size of the MC sample is 0.2\%. 
The uncertainty associated with the total number of $\psi(3686)$ events is 0.5\%~\cite{BESIII:2024lks}. 
The uncertainties arising from the knowledge of the branching fractions for $\Omega^- \to \Lambda K^-$ and $\Lambda \to p \pi^-$ are 1.0\% and 0.8\%~\cite{PDG}, respectively.

\hspace{0.7cm} The systematic uncertainties are summarized in Table~\ref{tab:systematic}. Assuming that all sources are independent, the total systematic uncertainty on the branching fraction of $\psi(3686) \to \Omega^- K^+\bar{\Xi}^0$
is determined to be 6.5\% by adding them in quadrature. 



\hspace{0.7cm}The signal significance is estimated to be $7.7\sigma$ after considering the systematic effects of the requirements of $M_{\Omega^-}$, and the signal and 
background shapes in the fit to $RM_{\Omega^- K^+}$. 

\begin{table}[htbp]
    \caption{Relative systematic uncertainties in the branching fraction measurement. }
    \label{tab:systematic}
    \begin{center}
    \begin{tabular} {l c c c}
        \hline \hline
        Source & Uncertainty $(\%)$ \\
        \hline
        Kaon tracking  & 2.0    \\
        Kaon PID    & 2.0    \\
        $\Lambda$ reconstruction & 4.1 \\
        Mass window of $M_{\Omega^-}$                 & 0.6    \\
        Signal shape     & 0.4    \\
        Background shape & 0.4   \\ 
        $f_\textrm{c}$ factor  & 0.3 \\
        MC generator & 3.9 \\
        MC sample size     & 0.2    \\
        $\mathcal{B}_{\Omega^- \to \Lambda K^-}$    & 1.0    \\
        $\mathcal{B}_{\Lambda \to p \pi^-}$           & 0.8    \\
        Number of $\psi(3686)$ events  & 0.5    \\
        \hline
        Total         & 6.5   \\
        \hline \hline
    \end{tabular}
    \end{center}
\end{table}

\section{Summary}
\label{sec:summary}
In summary, using the world’s largest $\psi(3686)$ sample taken with the BESIII detector, we observe the $\psi(3686) \to \Omega^- K^+\bar{\Xi}^0 +c.c.$ decay for the first time by employing a partial reconstruction method.
The measured branching fraction is 
$\mathcal{B}(\psi(3686) \to \Omega^- K^+\bar{\Xi}^0 +c.c.) = (2.78 \pm 0.40 \pm 0.18) \times 10^{-6}$, where the
first uncertainty is statistical and the second is systematic. 
This result provides useful information for understanding the dynamics of $\psi(3686)$ decays.
With the current sample size, we do not observe any clear evidence of possible hyperon excited states. 
A larger data sample would be helpful for studying the decay dynamics of this process and explore the presence of potential excited baryon states. 
\textbf{Acknowledgement}

The BESIII Collaboration thanks the staff of BEPCII and the IHEP computing center for their strong support. This work is supported in part by National Key R\&D Program of China under Contracts Nos. 2020YFA0406300, 2020YFA0406400; National Natural Science Foundation of China (NSFC) under Contracts Nos. 11635010, 11735014, 11835012, 11935015, 11935016, 11935018, 11961141012, 12025502, 12035009, 12035013, 12061131003, 12165022, 12192260, 12192261, 12192262, 12192263, 12192264, 12192265, 12221005, 12225509, 12235017, 12342502; the Chinese Academy of Sciences (CAS) Large-Scale Scientific Facility Program; the CAS Center for Excellence in Particle Physics (CCEPP); Joint Large-Scale Scientific Facility Funds of the NSFC and CAS under Contract No. U1832207; CAS Key Research Program of Frontier Sciences under Contracts Nos. QYZDJ-SSW-SLH003, QYZDJ-SSW-SLH040; 100 Talents Program of CAS; The Institute of Nuclear and Particle Physics (INPAC) and Shanghai Key Laboratory for Particle Physics and Cosmology; Yunnan Fundamental Research Project under Contract No. 202301AT070162; European Union's Horizon 2020 research and innovation programme under Marie Sklodowska-Curie grant agreement under Contract No. 894790; German Research Foundation DFG under Contracts Nos. 455635585, Collaborative Research Center CRC 1044, FOR5327, GRK 2149; Istituto Nazionale di Fisica Nucleare, Italy; Ministry of Development of Turkey under Contract No. DPT2006K-120470; National Research Foundation of Korea under Contract No. NRF-2022R1A2C1092335; National Science and Technology fund of Mongolia; National Science Research and Innovation Fund (NSRF) via the Program Management Unit for Human Resources \& Institutional Development, Research and Innovation of Thailand under Contract No. B16F640076; Polish National Science Centre under Contract No. 2019/35/O/ST2/02907; The Swedish Research Council; U. S. Department of Energy under Contract No. DE-FG02-05ER41374.

\bibliographystyle{JHEP}

\begin{thebibliography}{10}

\bibitem{E598:1974sol}
{\scshape E598} collaboration, \emph{{Experimental observation of a heavy
  particle $J$}},
  \href{https://doi.org/10.1103/PhysRevLett.33.1404}{\emph{Phys. Rev. Lett.}
  {\bfseries 33} (1974) 1404}.

\bibitem{BESIII:2024lks}
{\scshape BESIII} collaboration, \emph{{Determination of the number of $\psi(3686)$ events taken at BESIII}},
  [\href{https://arxiv.org/abs/2403.06766}{{\ttfamily arXiv:2403.06766}}].
  
\bibitem{PhysRevLett.33.1406}
J.~E. Augustin, A.~M. Boyarski, M.~Breidenbach, F.~Bulos, J.~T. Dakin, G.~J.
  Feldman et~al., \emph{Discovery of a narrow resonance in
  ${e}^{+}{e}^{\ensuremath{-}}$ annihilation},
  \href{https://doi.org/10.1103/PhysRevLett.33.1406}{\emph{Phys. Rev. Lett.}
  {\bfseries 33} (1974) 1406}.

\bibitem{Asner:2008nq}
D.~M. Asner et~al., \emph{{Physics at BES-III}}, {\emph{Int. J. Mod. Phys. A}
  {\bfseries 24} (2009) S1} [\href{https://arxiv.org/abs/0809.1869}{{\ttfamily
  arXiv:0809.1869}}].

\bibitem{BESIII:2020lkm}
{\scshape BESIII} collaboration, \emph{{Model-independent determination of the
  spin of the $\Omega^-$ and Its polarization alignment in $\psi(3686)\to
  \Omega^- \bar \Omega^+$}},
  \href{https://doi.org/10.1103/PhysRevLett.126.092002}{\emph{Phys. Rev. Lett.}
  {\bfseries 126} (2021) 092002}
  [\href{https://arxiv.org/abs/2007.03679}{{\ttfamily arXiv:2007.03679}}].

\bibitem{BESIII:2023olq}
{\scshape BESIII} collaboration, \emph{{Observation of the decay
  \ensuremath{\chi_{cJ}}\textrightarrow{}\ensuremath{\Omega^-}\ensuremath{\bar{\Omega}^+}}},
  \href{https://doi.org/10.1103/PhysRevD.107.092004}{\emph{Phys. Rev. D}
  {\bfseries 107} (2023) 092004}
  [\href{https://arxiv.org/abs/2302.12579}{{\ttfamily arXiv:2302.12579}}].

\bibitem{Rybicki:2009zza}
A.~Rybicki, \emph{{Hadronic interactions at the CERN SPS: resonance decays
  versus parton dynamics}},
  \href{https://doi.org/10.1142/S0217751X09043717}{\emph{Int. J. Mod. Phys. A}
  {\bfseries 24} (2009) 385}.

\bibitem{PDG}
{\scshape Particle Data Group} collaboration, \emph{{Review of Particle
  Physics}}, \href{https://doi.org/10.1093/ptep/ptac097}{\emph{PTEP} {\bfseries
  2022} (2022) 083C01}.

\bibitem{BES:2003aic}
{\scshape BES} collaboration, \emph{{Observation of a near threshold
  enhancement in th p anti-p mass spectrum from radiative \ensuremath{J/\psi}
  \textrightarrow{}\ensuremath{\gamma} \ensuremath{p\bar{p}} decays}},
  \href{https://doi.org/10.1103/PhysRevLett.91.022001}{\emph{Phys. Rev. Lett.}
  {\bfseries 91} (2003) 022001}
  [\href{https://arxiv.org/abs/hep-ex/0303006}{{\ttfamily hep-ex/0303006}}].

\bibitem{BES:2004fgd}
{\scshape BES} collaboration, \emph{{Observation of a threshold enhancement in
  the \ensuremath{p\bar{p}} invariant mass spectrum}},
  \href{https://doi.org/10.1103/PhysRevLett.93.112002}{\emph{Phys. Rev. Lett.}
  {\bfseries 93} (2004) 112002}
  [\href{https://arxiv.org/abs/hep-ex/0405050}{{\ttfamily hep-ex/0405050}}].

\bibitem{BESIII:2012koo}
{\scshape BESIII} collaboration, \emph{{Measurements of $\psi^\prime \to
  \bar{p} K^+ \Sigma^0$ and $\chi_{cJ} \to \bar{p} K^+ \Lambda$}},
  \href{https://doi.org/10.1103/PhysRevD.87.012007}{\emph{Phys. Rev. D}
  {\bfseries 87} (2013) 012007}
  [\href{https://arxiv.org/abs/1211.5631}{{\ttfamily arXiv:1211.5631}}].

\bibitem{Capstick:2000qj}
S.~Capstick and W.~Roberts, \emph{{Quark models of baryon masses and decays}},
  \href{https://doi.org/10.1016/S0146-6410(00)00109-5}{\emph{Prog. Part. Nucl.
  Phys.} {\bfseries 45} (2000) S241}
  [\href{https://arxiv.org/abs/nucl-th/0008028}{{\ttfamily nucl-th/0008028}}].

\bibitem{Aston:1987bb}
D.~Aston et~al., \emph{{Observation of $\Omega^{*-}$ production in $K^- p$
  interactions at 11 {GeV}/$c$}},
  \href{https://doi.org/10.1016/0370-2693(87)90238-3}{\emph{Phys. Lett. B}
  {\bfseries 194} (1987) 579}.

\bibitem{BESIII:2017tvm}
{\scshape BESIII} collaboration, \emph{{Determination of the number of
  $\psi(3686)$ events at BESIII}},
  \href{https://doi.org/10.1088/1674-1137/42/2/023001}{\emph{Chin. Phys. C}
  {\bfseries 42} (2018) 023001}
  [\href{https://arxiv.org/abs/1709.03653}{{\ttfamily arXiv:1709.03653}}],
  {With the same method, the total number of $\psi(3686)$ events collected in
  2009, 2012 and 2021 is determined to be $27.13 \times 10^8$ with an
  uncertainty of 0.5\% as a preliminary result.}

\bibitem{BESIII:2009fln}
{\scshape BESIII} collaboration, \emph{{Design and construction of the BESIII
  detector}}, \href{https://doi.org/10.1016/j.nima.2009.12.050}{\emph{Nucl.
  Instrum. Meth. A} {\bfseries 614} (2010) 345}
  [\href{https://arxiv.org/abs/0911.4960}{{\ttfamily arXiv:0911.4960}}].

\bibitem{BESIII:2020nme}
{\scshape BESIII} collaboration, \emph{{Future physics programme of BESIII}},
  \href{https://doi.org/10.1088/1674-1137/44/4/040001}{\emph{Chin. Phys. C}
  {\bfseries 44} (2020) 040001}
  [\href{https://arxiv.org/abs/1912.05983}{{\ttfamily arXiv:1912.05983}}].

\bibitem{Yu:2016cof}
C.~Yu et~al., \emph{{BEPCII performance and beam dynamics studies on
  luminosity}},  in \emph{{7th International Particle Accelerator Conference}},
  p.~TUYA01, 2016,
  \href{https://doi.org/doi:10.18429/JACoW-IPAC2016-TUYA01}{DOI}.

\bibitem{Huang:2022wuo}
K.~X. Huang, Z.~J. Li, Z.~Qian, J.~Zhu, H.~Y. Li, Y.~M. Zhang et~al.,
  \emph{{Method for detector description transformation to unity and
  application in BESIII}},
  \href{https://doi.org/10.1007/s41365-022-01133-8}{\emph{Nucl. Sci. Tech.}
  {\bfseries 33} (2022) 142}
  [\href{https://arxiv.org/abs/2206.10117}{{\ttfamily arXiv:2206.10117}}].

\bibitem{Li:2017eToF}
X.~Li et~al., \emph{{Study of MRPC technology for BESIII endcap-TOF upgrade}},
  \href{https://doi.org/10.1007/s41605-017-0014-2}{\emph{Radiat. Detect.
  Technol. Methods} {\bfseries 1} (2017) 13}.

\bibitem{Guo:2017eToF}
Y.~X. Guo et~al., \emph{{The study of time calibration for upgraded end cap TOF
  of BESIII}}, \href{https://doi.org/10.1007/s41605-017-0012-4}{\emph{Radiat.
  Detect. Technol. Methods} {\bfseries 1} (2017) 15}.

\bibitem{Cao:2020ibk}
P.~Cao et~al., \emph{{Design and construction of the new BESIII endcap
  Time-of-Flight system with MRPC Technology}},
  \href{https://doi.org/10.1016/j.nima.2019.163053}{\emph{Nucl. Instrum. Meth.
  A} {\bfseries 953} (2020) 163053}.

\bibitem{GEANT4:2002zbu}
{\scshape GEANT4} collaboration, \emph{{GEANT4--a simulation toolkit}},
  \href{https://doi.org/10.1016/S0168-9002(03)01368-8}{\emph{Nucl. Instrum.
  Meth. A} {\bfseries 506} (2003) 250}.

\bibitem{Jadach:2000ir}
S.~Jadach, B.~F.~L. Ward and Z.~Was, \emph{{Coherent exclusive exponentiation
  for precision Monte Carlo calculations}},
  \href{https://doi.org/10.1103/PhysRevD.63.113009}{\emph{Phys. Rev. D}
  {\bfseries 63} (2001) 113009}
  [\href{https://arxiv.org/abs/hep-ph/0006359}{{\ttfamily hep-ph/0006359}}].

\bibitem{Lange:2001uf}
D.~J. Lange, \emph{{The EvtGen particle decay simulation package}},
  \href{https://doi.org/10.1016/S0168-9002(01)00089-4}{\emph{Nucl. Instrum.
  Meth. A} {\bfseries 462} (2001) 152}.

\bibitem{Ping:2008zz}
R.~G. Ping, \emph{{Event generators at BESIII}},
  \href{https://doi.org/10.1088/1674-1137/32/8/001}{\emph{Chin. Phys. C}
  {\bfseries 32} (2008) 599}.

\bibitem{Chen:2000tv}
J.~C. Chen, G.~S. Huang, X.~R. Qi, D.~H. Zhang and Y.~S. Zhu, \emph{Event
  generator for \ensuremath{J/\psi} and \ensuremath{\psi(3686)} decay},
  \href{https://doi.org/10.1103/PhysRevD.62.034003}{\emph{Phys.Rev.D}
  {\bfseries 62} (2000) 034003}.

\bibitem{Yang:2014vra}
R.~L. Yang, R.~G. Ping and H.~Chen, \emph{Tuning and validation of the
  lundcharm model with \ensuremath{J/\psi} decays},
  \href{https://doi.org/10.1088/0256-307X/31/6/061301}{\emph{Chin.Phys.Lett.}
  {\bfseries 31} (2014) 061301}.

\bibitem{Richter-Was:1992hxq}
E.~Richter-Was, \emph{{QED bremsstrahlung in semileptonic $B$ and leptonic tau
  decays}}, \href{https://doi.org/10.1016/0370-2693(93)90062-M}{\emph{Phys.
  Lett. B} {\bfseries 303} (1993) 163}.

\bibitem{Xu:2009zzg}
M.~Xu et~al., \emph{{Decay vertex reconstruction and 3-dimensional lifetime
  determination at BESIII}},
  \href{https://doi.org/10.1088/1674-1137/33/6/005}{\emph{Chin. Phys. C}
  {\bfseries 33} (2009) 428}.

\bibitem{Zhou:2020ksj}
X.~Zhou, S.~Du, G.~Li and C.~Shen, \emph{{TopoAna: A generic tool for the event
  type analysis of inclusive Monte-Carlo samples in high energy physics
  experiments}}, \href{https://doi.org/10.1016/j.cpc.2020.107540}{\emph{Comput.
  Phys. Commun.} {\bfseries 258} (2021) 107540}
  [\href{https://arxiv.org/abs/2001.04016}{{\ttfamily arXiv:2001.04016}}].

\bibitem{Ke:2023qzc}
B.-C. Ke, J.~Koponen, H.-B. Li and Y.~Zheng, \emph{{Recent Progress in Leptonic
  and Semileptonic Decays of Charmed Hadrons}},
  \href{https://doi.org/10.1146/annurev-nucl-110222-044046}{\emph{Ann. Rev.
  Nucl. Part. Sci.} {\bfseries 73} (2023) 285}
  [\href{https://arxiv.org/abs/2310.05228}{{\ttfamily arXiv:2310.05228}}].

\bibitem{Ablikim:2013ntc}
{\scshape BESIII} collaboration, \emph{{Measurement of the integrated
  luminosities of the data taken by BESIII at $\sqrt{s}=$3.650 and 3.773 GeV}},
  \href{https://doi.org/10.1088/1674-1137/37/12/123001}{\emph{Chin. Phys. C}
  {\bfseries 37} (2013) 123001}
  [\href{https://arxiv.org/abs/1307.2022}{{\ttfamily arXiv:1307.2022}}], {The
  integrated luminosity of the $\psi(3773)$ data samples collected in 2010,
  2011 and 2022 is determined to be $(7.93\pm0.02)~\textrm{fb}^{-1}$ as a
  preliminary result.}

\bibitem{BESIII:2018ldc}
{\scshape BESIII} collaboration, \emph{{Measurement of $e^{+} e^{-} \rightarrow
  K^{+} K^{-}$ cross section at $\sqrt{s} = 2.00 - 3.08$ GeV}},
  \href{https://doi.org/10.1103/PhysRevD.99.032001}{\emph{Phys. Rev. D}
  {\bfseries 99} (2019) 032001}
  [\href{https://arxiv.org/abs/1811.08742}{{\ttfamily arXiv:1811.08742}}].

\bibitem{BESIII:2023drj}
{\scshape BESIII} collaboration, \emph{{Tests of $CP$ symmetry in entangled
  \ensuremath{\Xi^{0}} - \ensuremath{\bar{\Xi}^{0}} pairs}},
  \href{https://doi.org/10.1103/PhysRevD.108.L031106}{\emph{Phys. Rev. D}
  {\bfseries 108} (2023) L031106}
  [\href{https://arxiv.org/abs/2305.09218}{{\ttfamily arXiv:2305.09218}}].

\bibitem{BESIII:2023ldd}
{\scshape BESIII} collaboration, \emph{{Measurements of the absolute branching
  fractions of $\Omega^-$ decays and test of the $\Delta I = 1/2$ rule}},
  \href{https://arxiv.org/abs/2309.06368}{{\ttfamily arXiv:2309.06368}}.

\end{thebibliography}

\providecommand{\href}[2]{#2}\begingroup\raggedright\endgroup

\end{document}